\documentclass[prb,reprint,showpacs,floatfix,superscriptaddress]{revtex4-1}
\usepackage[pdftex]{graphicx} 
\usepackage{bm} 
\usepackage{amssymb} 
\usepackage{amsmath} 
\usepackage{hyperref}
\usepackage{booktabs}
\usepackage{multirow}
\usepackage{siunitx}
\usepackage{footmisc}
\usepackage{csquotes}
\usepackage{booktabs}
\usepackage[latin1]{inputenc}
\begin{document}
\title{Comparison of local density functionals based on electron gas and finite systems}
\date{\today}
\author{M.\ T.\ Entwistle}
\affiliation{Department of Physics, University of York, and European Theoretical Spectroscopy Facility, Heslington, York YO10 5DD, United Kingdom}
\author{M. Casula}
\affiliation{Institut de Min\'{e}ralogie, de Physique des Mat\'{e}riaux et de Cosmochimie (IMPMC), Sorbonne Universit\'{e}, CNRS UMR 7590, IRD UMR 206, MNHN, 4 Place Jussieu, 75252 Paris, France}
\author{R.\ W.\ Godby}
\affiliation{Department of Physics, University of York, and European Theoretical Spectroscopy Facility, Heslington, York YO10 5DD, United Kingdom}

\begin{abstract}   
A widely used approximation to the exchange-correlation functional in density functional theory is the local density approximation (LDA), typically derived from the properties of the homogeneous electron gas (HEG). We previously introduced a set of alternative LDAs constructed from one-dimensional systems of one, two, and three electrons that resemble the HEG within a finite region. We now construct a HEG-based LDA appropriate for spinless electrons in one dimension and find that it is remarkably similar to the finite LDAs. As expected, all LDAs are inadequate in low-density systems where correlation is strong. However, exploring the small but significant differences between the functionals, we find that the finite LDAs give better densities and energies in high-density exchange-dominated systems, arising partly from a better description of the self-interaction correction. 
\end{abstract}

\maketitle

\section{Introduction} 
Density functional theory \cite{DFT} (DFT) is the most popular method to calculate the ground-state properties of many-electron systems \cite{LDA_Success,LDA_Success2,DFT2,DFT3,DFT4,DFT5}. In the widely employed Kohn-Sham \cite{KS_LDA} (KS) formalism of DFT, the real system of interacting electrons is mapped onto a fictitious system of noninteracting electrons moving in an effective local potential, with both systems having the same electron density. While in principle an exact theory, in practice the accuracy of DFT calculations is constrained by our ability to approximate the exchange-correlation (xc) part of the KS functional, whose exact form is unknown. Identifying properties of the exact xc functional that are missing in commonly used approximations is vital for further developments.

A widely used approximation is the local density approximation \cite{KS_LDA} (LDA) which assumes that the true xc functional is solely dependent on the electron density at each point in the system. LDAs are traditionally derived from knowledge of the xc energy of the homogeneous electron gas \cite{QMC_HEG} (HEG), a model system where the exchange energy \footnote{Throughout this paper, we take the exchange energy to be the exchange energy of a self-consistent Hartree-Fock calculation.} is known analytically and the correlation energy \footnote{Throughout this paper, we take the correlation energy to be the difference between the exact energy of the many-electron system and the energy of a self-consistent Hartree-Fock calculation.} is usually calculated using quantum Monte Carlo simulations. LDAs have been hugely successful in many cases \cite{LDA_Success,LDA_Success2}, however, their validity breaks down in a number of important situations \cite{ALDA,ALDA2,ALDA3,ALDA4,ALDA5,ALDA6,ALDA7,ALDA8,ALDA9}, particularly when there is strong correlation. They are known to miss out some critical features that are present in the exact xc potential, such as the cancellation of the spurious electron self-interaction \cite{PerdewZunger,SIC,SIC2}, or the Coulomb-type $-1/r$ decay of the xc potential far from a finite system \cite{CoulombDecay,CoulombDecay/ExpDecay}, instead following an incorrect exponential decay \cite{PerdewZunger,CoulombDecay/ExpDecay}. They also fail to capture the derivative discontinuity \cite{DD1,DD2,DD3}, the discontinuous nature of the derivative of the xc energy with respect to electron number $N$, at integer $N$.

In a previous paper \cite{Entwistle_LDA}, we introduced a set of LDAs which, in contrast to the traditional HEG LDA, were constructed from systems of one, two, and three electrons which resembled the HEG within a \textit{finite} region. Illustrating our approach in one dimension (1D), we found that the three LDAs were remarkably similar to one another. In this paper, we construct a 1D HEG LDA through suitable diffusion Monte Carlo \cite{Michele_QMC} (DMC) techniques, along with a revised set of LDAs constructed from finite systems. We compare the finite and HEG LDAs with one another to demonstrate that local approximations constructed from finite systems are a viable alternative, and explore the nature of any differences between them.

In order to test the LDAs, we employ our iDEA code \cite{iDEA} which solves the many-electron Schr\"{o}dinger equation exactly for model finite systems to determine the exact, fully-correlated, many-electron wave function. Using this to obtain the \textit{exact electron density}, we then utilize our reverse engineering algorithm to find the exact KS system. In our calculations we use \textit{spinless} electrons to more closely approach the nature of exchange and correlation in many-electron systems,\footnote{Spinless electrons obey the Pauli principle but are restricted to a single spin type. Systems of two or three spinless electrons exhibit features that would need a larger number of spin-half electrons to become apparent. For example, two spinless electrons experience the exchange effect, which is not the case for two spin-half electrons in an $S = 0$ state. Furthermore, spinless KS electrons occupy a greater number of KS orbitals.} which interact via the appropriately softened Coulomb repulsion \cite{SoftenedCoulomb}\textsuperscript{,} \footnote{We use Hartree atomic units: $m_{\mathrm{e}} = \hbar = e = 4\pi \varepsilon_{0} = 1$} $(|x-x'|+1)^{-1}$.

\section{Set of LDAs}

\subsection{LDAs from finite systems}
\begin{figure}[htbp]      
\centering
\includegraphics[width=1.0\linewidth]{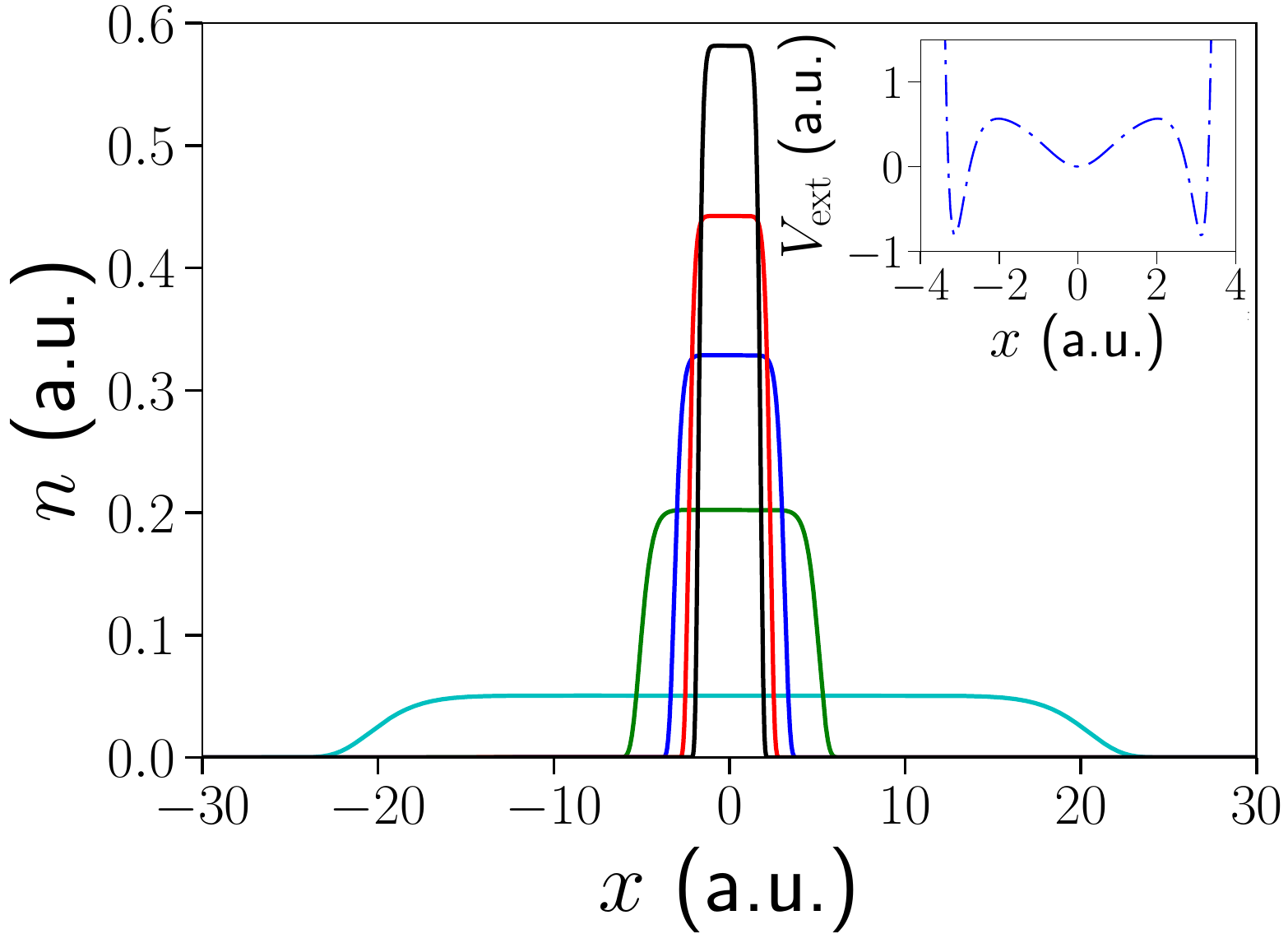}
\caption{The exact many-body electron density (solid lines) for a selection of the two-electron slab systems. The density is locally homogeneous across a plateau region and decays exponentially at the edges. Inset: the external potential for a typical two-electron slab system (middle density in main figure).}
\label{slabs}
\end{figure}

In Ref.~\onlinecite{Entwistle_LDA} we chose a set of finite locally homogeneous systems in order to mimic the HEG, which we referred to as \enquote{slabs} (Fig.~\ref{slabs}). We generated sets of one-electron ($1e$), two-electron ($2e$), and three-electron ($3e$) slab systems over a typical density range (up to 0.6 a.u.) and in each case calculated the exact xc energy $E_{\mathrm{xc}}$. From this we parametrized the xc energy density $\varepsilon_{\mathrm{xc}} = E_{\mathrm{xc}}/N$ in terms of the electron density of the plateau region of the slabs, repeating for the $1e$, $2e$, and $3e$ set.

To approximate the xc energy of an inhomogeneous system, the LDA focuses on the local electron density at each point in the system:
\begin{equation}
E_{\mathrm{xc}}^{\mathrm{LDA}}[n] = \int n(x)\varepsilon_{\mathrm{xc}}(n) \ dx \label{E_xc},
\end{equation}
where in a conventional LDA $\varepsilon_{\mathrm{xc}}(n)$ is the xc energy density of a HEG of density $n$. This approximation becomes exact in the limit of the HEG, and so it is a reasonable requirement for the finite LDAs to become exact in the limit of the slab systems. Due to the initial parametrization of $\varepsilon_{\mathrm{xc}}(n)$ focusing on the plateau regions of the slabs (i.e., ignoring the inhomogeneous regions at the edges), we used a \textit{refinement process} \cite{Entwistle_LDA} in order to fulfill this requirement.

The refined form for the xc energy density in the three finite LDAs has now been increased from the four-parameter fit in Ref.~\onlinecite{Entwistle_LDA} to a seven-parameter fit\footnote{We have significantly increased the precision of the calculations for the slab systems in order to do this. The numerical difference between the new seven-parameter fits and original four-parameter fits is less than 1$\%$ in $\varepsilon_{\mathrm{xc}}$ across the density range used in constructing the LDAs (except in the very low-density region $n< 0.06$ a.u.). This has allowed us to resolve the differences between the four LDAs in fine detail.} in this paper:
\begin{equation} 
\varepsilon_{\mathrm{xc}}(n) = (A + B n + C n^{2} + D n^{3} + E n^{4} + F n^{5}) n^{G}, \label{exc_finite}
\end{equation}
where the optimal parameters for each LDA are given in Table~\ref{finite_table}. The xc potential $V_{\mathrm{xc}}$ is defined as the functional derivative of the xc energy which in the LDA reduces to a simple form \footnote{See Supplemental Material for the parametric form of the xc potential in the finite LDAs.}:
\begin{equation}
V_{\mathrm{xc}}^{\mathrm{LDA}}(n) = \varepsilon_{\mathrm{xc}}(n(x)) + n(x)\frac{d\varepsilon_{\mathrm{xc}}}{dn}\bigg|_{n(x)}. \label{V_xc} 
\end{equation}

\newcommand\T{\rule{0pt}{2.6ex}}       
\newcommand\B{\rule[-1.2ex]{0pt}{0pt}} 
\begin{table}[htb]
\caption{Optimal fit parameters for $\varepsilon_{\mathrm{xc}}(n)$ in the finite LDAs. The last two rows contain the mean absolute error (MAE) and root-mean-square error (RMSE) of the fits. $\varepsilon_{\mathrm{xc}}(n)$ is graphed in Sec.~\ref{comparison_ldas} below.}
\label{finite_table}
\resizebox{\columnwidth}{!}{%
\begin{ruledtabular}
\begin{tabular}{l c c c}
   Parameter & $1e$ value               & $2e$ value               & $3e$ value                \B \\
   \hline
   $A$       & $-1.2202$                  & $-1.0831$                  & $-1.1002$                 \T \\
   $B$       & 3.6838                   & 2.7609                   & 2.9750                     \\
   $C$       & $-11.254$                  & $-7.1577$                  & $-8.1618$                    \\
   $D$       & 23.169                   & 12.713                   & 15.169                     \\
   $E$       & $-26.299$                  & $-12.755$                  & $-15.776$                    \\
   $F$       & 12.282                   & 5.3817                   & 6.8494                     \\
   $G$       & 7.4876 $\times 10^{-1}$  & 7.0955 $\times 10^{-1}$  & 7.0907 $\times 10^{-1}$ \B \\

   MAE       & 1.3 $\times 10^{-4}$     & 1.2 $\times 10^{-4}$     & 9.9 $\times 10^{-5}$ \T \\
   RMSE      & 1.9 $\times 10^{-3}$     & 5.1 $\times 10^{-4}$     & 3.8 $\times 10^{-4}$    
\end{tabular}
\end{ruledtabular}
}
\end{table}

\subsection{HEG exchange functional}
In Ref.~\onlinecite{Entwistle_LDA} we solved the Hartree-Fock equations to find the exact exchange energy density $\varepsilon_{\mathrm{x}}$ for a fully spin-polarized [$\zeta = 1$ where $\zeta \equiv (N^{\uparrow}-N^{\downarrow})/N$] 1D HEG of density $n$ consisting of an infinite number of electrons interacting via the softened Coulomb repulsion $u(x-x') = (|x-x'|+1)^{-1}$:
\begin{equation}
\varepsilon_{\mathrm{x}}(n) = -\frac{1}{8\pi^{2}n}\int_{-\pi n}^{\pi n}dk \int_{-\pi n}^{\pi n}dk' \ u(k'-k),\label{HF}
\end{equation}
where the Fourier transform of $u(x-x')$ is integrated over the plane defined by the Fermi wave vector $k_{\mathrm{F}}=\pi n$.

Solving Eq.~(\ref{HF}) for the range of densities we used in the finite LDAs, we parametrized $\varepsilon_{\mathrm{x}}(n)$. Once again, we have increased our fit from four parameters to seven parameters, as in Eq.~(\ref{exc_finite}) above \footnote{See Supplemental Material for the parametric form of the exchange potential in the HEG LDA.}. The optimal parameters are given in Table~\ref{hegx_table}. The $\varepsilon_{\mathrm{x}}(n)$ curve is shown in the inset of Fig.~\ref{ex_ec_heg}.

\begin{table}[htb]
\caption{Optimal fit parameters for $\varepsilon_{\mathrm{x}}(n)$ in the HEG LDA. The last two rows contain the mean absolute error (MAE) and root-mean-square error (RMSE) of the fit.}
\label{hegx_table}
\resizebox{\columnwidth}{!}{%
\begin{ruledtabular}
\begin{tabular}{l c}
   Parameter & Value                    \B \\
   \hline
   $A$       & $-1.1511$                  \T \\
   $B$       & 3.3440                      \\
   $C$       & $-9.7079$                     \\
   $D$       & 19.088                      \\
   $E$       & $-20.896$                     \\
   $F$       & 9.4861                      \\
   $G$       & 7.3586 $\times 10^{-1}$  \B \\

   MAE       & 6.5 $\times 10^{-5}$      \T \\
   RMSE      & 7.2 $\times 10^{-4}$   
\end{tabular}
\end{ruledtabular}
}
\end{table}

\subsection{HEG correlation functional}
We use the lattice regularized diffusion Monte Carlo (LRDMC) algorithm \cite{Michele_QMC} to compute the ground-state energy of the fully spin-polarized HEG over a wide range of densities, much higher than the 0.6 a.u. limit used in the finite LDAs. This is in order to ensure the resultant parametrization of the correlation energy density $\varepsilon_{\mathrm{c}}$ reduces to the known high-density and low-density limits. We determine $\varepsilon_{\mathrm{c}}$ by subtracting the kinetic energy and $\varepsilon_{\mathrm{x}}$ contributions from the total energy.

\begin{figure}[htbp]      
\centering
\includegraphics[width=1.0\linewidth]{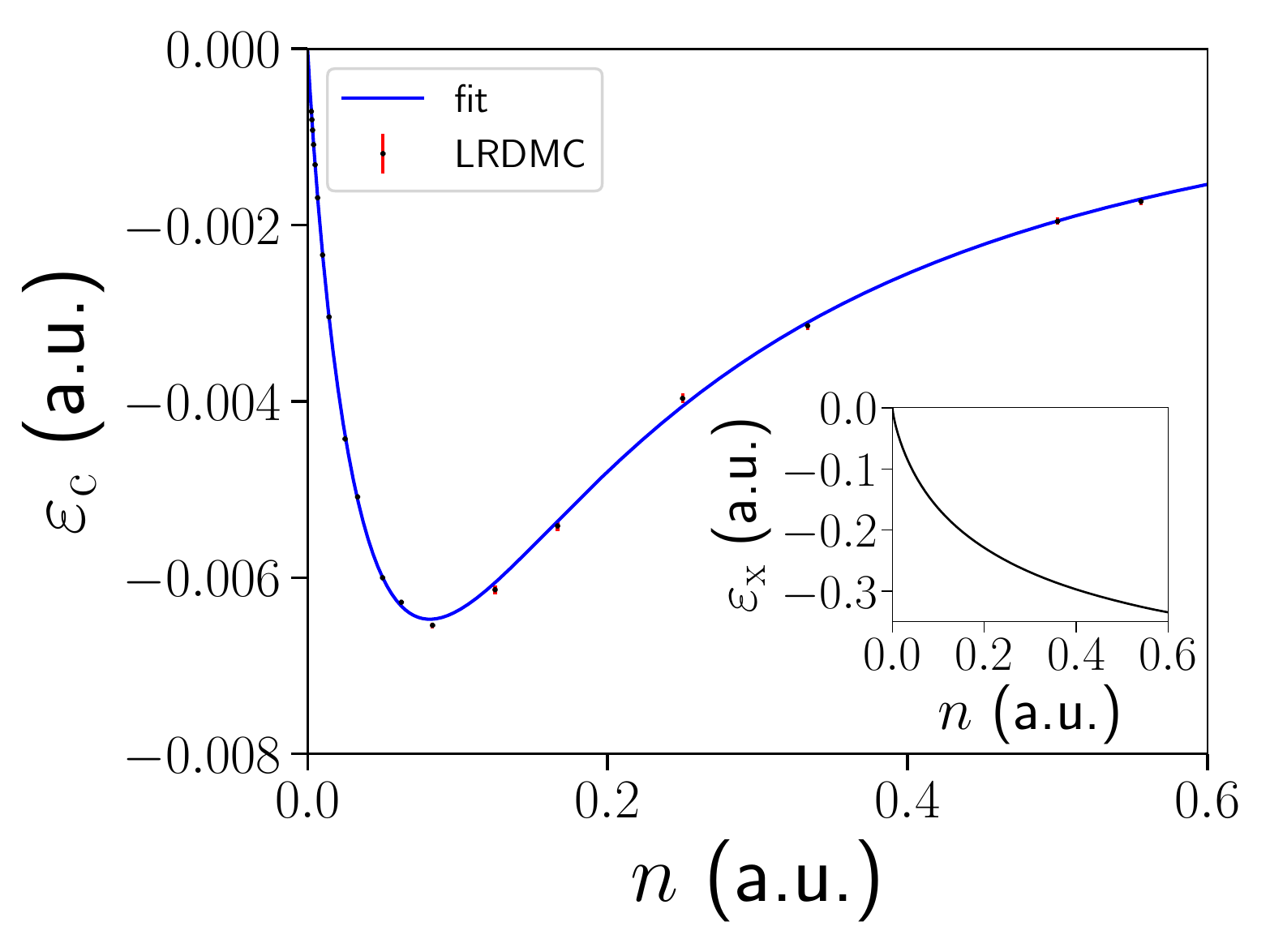}
\caption{The $\varepsilon_{\mathrm{c}}$ (with associated error bars) for a set of HEGs over the density range used in the finite LDAs. The fit applied (solid blue) becomes exact in the known high-density and low-density limits. Inset: The $\varepsilon_{\mathrm{x}}$ curve in the HEG LDA.}
\label{ex_ec_heg}
\end{figure}

To parametrize the correlation energy density we use a fit of the form \footnote{See Supplemental Material for the parametric form of the correlation potential in the HEG LDA.}:
\begin{equation}
\varepsilon_{\mathrm{c}}(r_{\mathrm{s}}) = -\frac{A_{\mathrm{RPA}} r_{\mathrm{s}} + E r_{\mathrm{s}}^{2}}{1 + B r_{\mathrm{s}} + C r_{\mathrm{s}}^{2} + D r_{\mathrm{s}}^{3}} \frac{\ln(1 + \alpha r_{\mathrm{s}} + \beta r_{\mathrm{s}}^{2})}{\alpha}, \label{ec_fit}
\end{equation}
where $r_{\mathrm{s}}$ is the Wigner-Seitz radius and is related to the density (in 1D) by $2 r_{\mathrm{s}} = 1/n$. The optimal parameters (with estimated errors) are given in Table~\ref{hegc_table}. The fit applied to the data is shown in Fig.~\ref{ex_ec_heg}.

\begin{table}[htb]
\caption{Optimal fit parameters with estimated errors in parentheses for $\varepsilon_{\mathrm{c}}(r_{\mathrm{s}})$ in the HEG LDA. The last two rows contain the mean absolute error (MAE) and root-mean-square error (RMSE) of the fit. Note: $A_{\mathrm{RPA}}$ has been determined from the high-density limit for $\varepsilon_{\mathrm{c}}$ (in which the random phase approximation (RPA) is exact \cite{RPA1,RPA2}), which is exactly fulfilled by our fit, and hence has no associated error.}
\label{hegc_table}
\resizebox{\columnwidth}{!}{%
\begin{ruledtabular}
\begin{tabular}{l c}
   Parameter           & Value                       \B \\
   \hline
   $A_{\mathrm{RPA}}$  & 9.415195 $\times 10^{-4}$   \T \\
   $B$                 & 2.601(5) $\times 10^{-1}$      \\
   $C$                 & 6.404(7) $\times 10^{-2}$      \\
   $D$                 & 2.48(3) $\times 10^{-4}$       \\
   $E$                 & 2.61(3) $\times 10^{-6}$       \\
   $\alpha$            & 1.254(2)                       \\
   $\beta$             & 28.8(1)                     \B \\

   MAE                 & 2.4 $\times 10^{-5}$        \T \\
   RMSE                & 1.3 $\times 10^{-4}$         
\end{tabular}
\end{ruledtabular}
}
\end{table}

The high-density limit (infinitely-weak correlation) of the parametrization is: 
\begin{equation}
\varepsilon_{\mathrm{c}}(r_{\mathrm{s}} \rightarrow 0) = -A_{\mathrm{RPA}}r_{\mathrm{s}}^{2},
\end{equation}
and its low-density limit (infinitely-strong correlation) is:
\begin{equation}
\varepsilon_{\mathrm{c}}(r_{\mathrm{s}} \rightarrow \infty) = -\frac{2E}{\alpha D}\frac{\ln(r_{\mathrm{s}})}{r_{\mathrm{s}}}.
\end{equation}
Therefore, the parametric form in Eq.~(\ref{ec_fit}) correctly reproduces the expected behavior of the correlation energy density in the high-density limit \cite{RPA1,RPA2} [$\varepsilon_{\mathrm{c}} \propto r_{\mathrm{s}}^{2}$] and low-density limit [$\varepsilon_{\mathrm{c}} \propto \ln(r_{\mathrm{s}})/r_{\mathrm{s}}$].

\subsection{Comparison of 1\textit{e}, 2\textit{e}, 3\textit{e} and HEG LDAs} \label{comparison_ldas}
\begin{figure}[htbp] 
\centering
\includegraphics[width=1.0\linewidth]{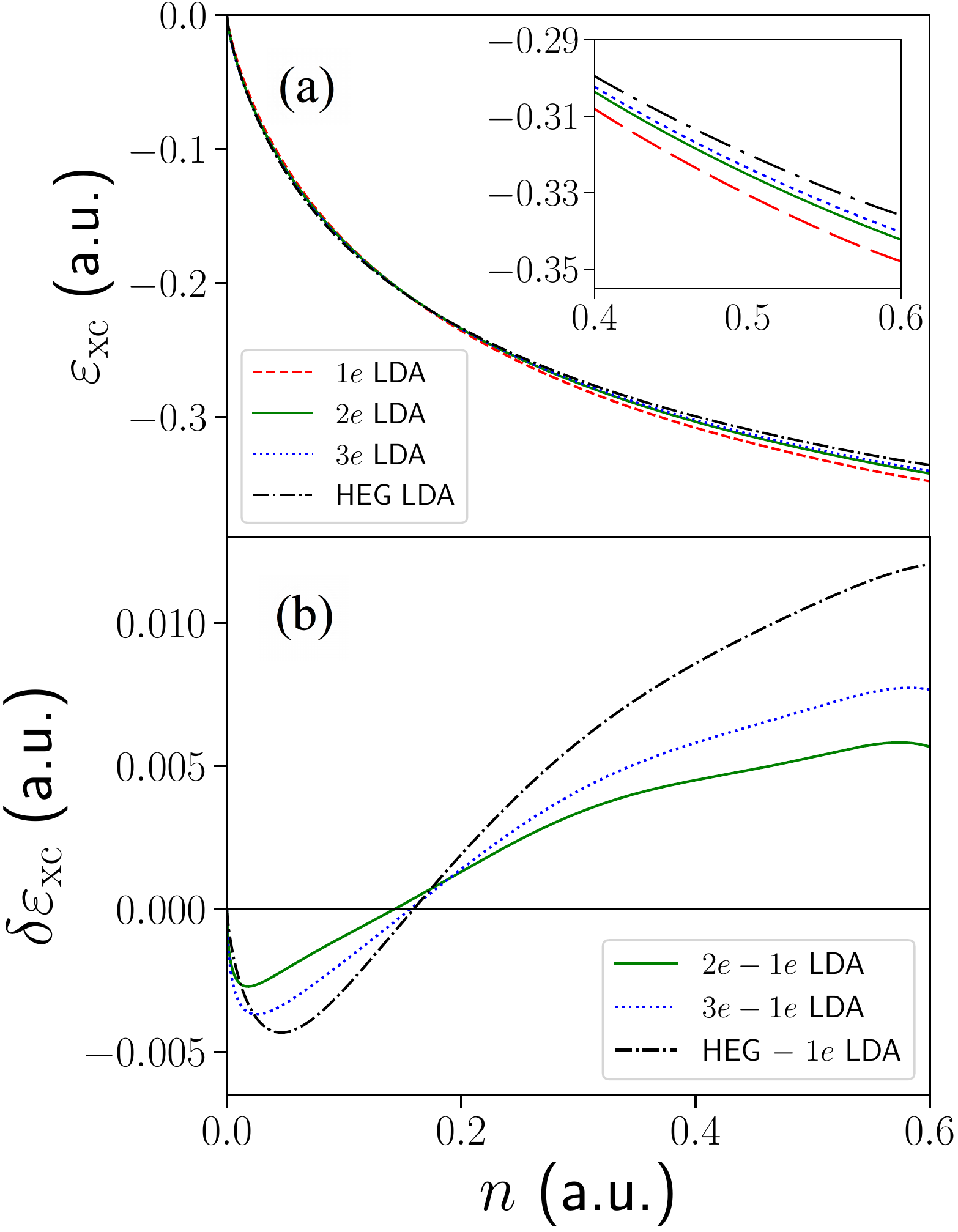}
\caption{(a) The $\varepsilon_{\mathrm{xc}}$ curves in the $1e$ (dashed red), $2e$ (solid green), $3e$ (dotted blue) and HEG (dotted-dashed black) LDAs. Inset: Close-up of the four curves at higher densities. The similarity between them is striking, with a clear progression from $1e$ to $2e$ to $3e$ to HEG. (b) The $1e$ LDA is used as a reference here. Plotted is its difference ($\delta \varepsilon_{\mathrm{xc}} = \varepsilon_{\mathrm{xc}} - \varepsilon_{\mathrm{xc}}^{1e}$) with the $2e$ (solid green), $3e$ (dotted blue) and HEG (dotted-dashed black) LDAs.}
\label{exc_combined}
\end{figure}

\begin{figure}[htbp] 
\centering
\includegraphics[width=1.0\linewidth]{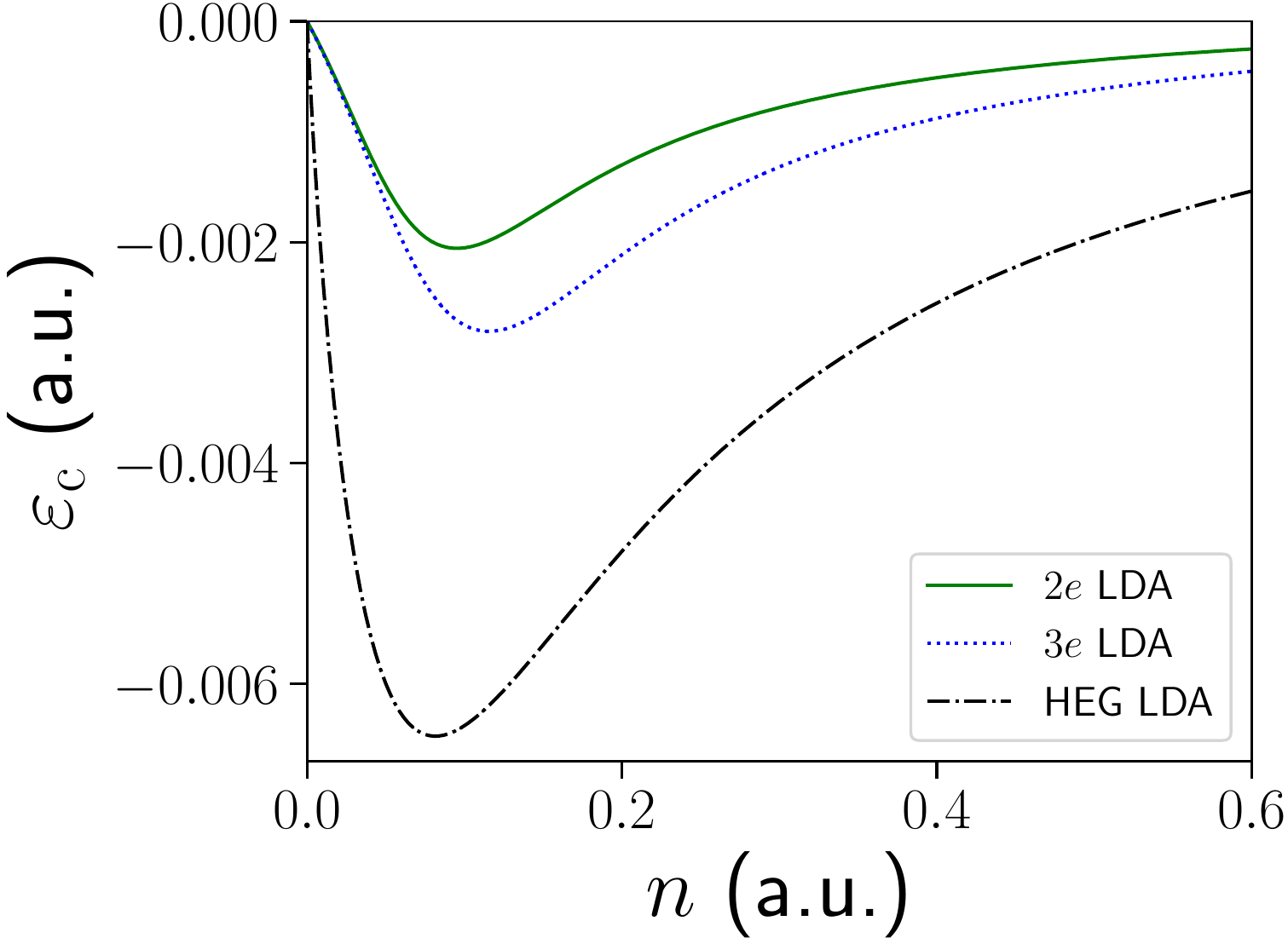}
\caption{We calculate the exact $\varepsilon_{\mathrm{c}}$ for the $2e$ (solid green line) and $3e$ (dotted blue line) slab systems through Hartree-Fock calculations. We plot these against the $\varepsilon_{\mathrm{c}}$ curve in the HEG LDA (dotted-dashed black line). The $\varepsilon_{\mathrm{c}}$ in the HEG LDA is much larger ($\sim$2--3 that of the $3e$ LDA and $\sim$3--4 that of the $2e$ LDA). While not a perfect comparison due to the refinement process used in the construction of the finite LDAs, it gives a useful indication of the size of $\varepsilon_{\mathrm{c}}$ in their $\varepsilon_{\mathrm{xc}}$ curves.}
\label{ec_comparison}
\end{figure}

\begin{figure}[htbp] 
\centering
\includegraphics[width=1.0\linewidth]{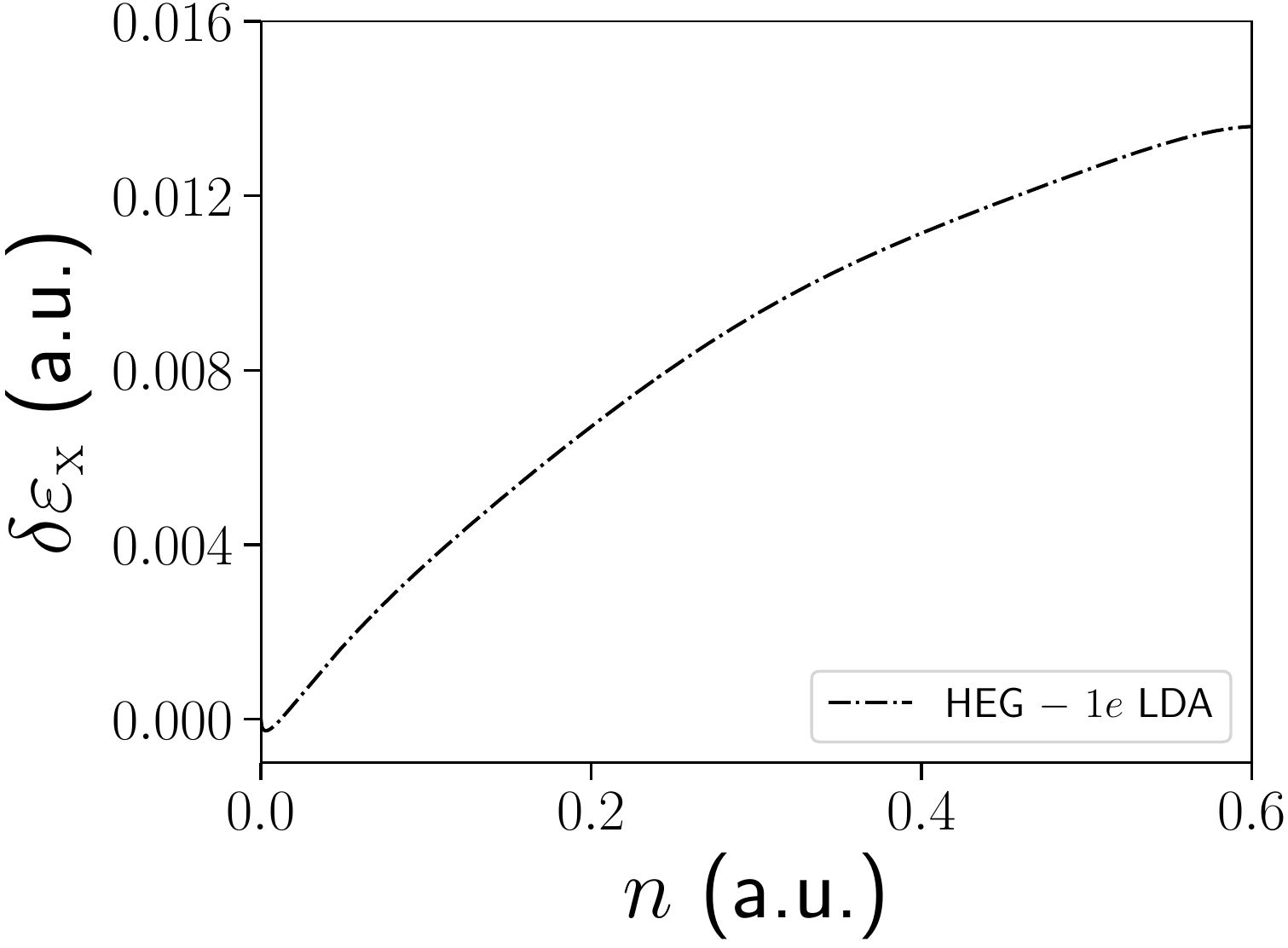}
\caption{The $\varepsilon_{\mathrm{x}}$ curve in the $1e$ LDA ($\varepsilon_{\mathrm{x}} = \varepsilon_{\mathrm{xc}}$) is used as a reference here. Plotted is its difference ($\delta \varepsilon_{\mathrm{x}} = \varepsilon_{\mathrm{x}} - \varepsilon_{\mathrm{x}}^{1e}$) with the $\varepsilon_{\mathrm{x}}$ curve in the HEG LDA ($\varepsilon_{\mathrm{x}} = \varepsilon_{\mathrm{xc}} - \varepsilon_{\mathrm{c}}$). It can be seen that the $1e$ LDA yields a larger exchange energy density than the HEG LDA at all densities. Note: This is not true in the very low-density region ($n < 0.012$), which we attribute to errors in the fits.}
\label{ex_comparison}
\end{figure}

Summing together the HEG exchange and correlation parametric fits, we can now compare the HEG LDA that we have developed against the three finite LDAs. The striking similarity between the four $\varepsilon_{\mathrm{xc}}$ curves can be seen in Fig.~\ref{exc_combined}(a). While very similar in the low-density range, there are some differences between them. These are highlighted in Fig.~\ref{exc_combined}(b) which, using the $1e$ LDA as a reference, plots its difference with the remaining LDAs. There is a competing balance between exchange and correlation. At low densities, these differences can be mainly attributed to $\varepsilon_{\mathrm{c}}$, which is entirely absent in the $1e$ LDA, and increases in magnitude as we progress to $2e$ to $3e$ to HEG (Fig.~\ref{ec_comparison}). As we move to higher densities in which the magnitude of $\varepsilon_{\mathrm{c}}$ decreases, and the magnitude of $\varepsilon_{\mathrm{x}}$ increases, the order of the four $\varepsilon_{\mathrm{xc}}$ curves reverses. They increasingly separate as we move to higher densities with the $1e$ LDA, which consists entirely of self-interaction correction, giving the largest magnitude for $\varepsilon_{\mathrm{xc}}$. By plotting the difference between the $1e$ LDA (where correlation is absent) and the exchange part of the HEG LDA (i.e., removing the correlation term), it can be seen that the $1e$ LDA yields a larger exchange energy density than the HEG LDA at all densities (Fig.~\ref{ex_comparison}).

The refinement process used in the construction of the finite LDAs focused on giving the correct $E_{\mathrm{xc}}$ in the limit of the slab systems, but did not ensure that the correct $V_{\mathrm{xc}}$, and by extension electron density, were reproduced (a property of HEG LDAs). We find that the finite LDAs are completely inadequate at reproducing the densities of the slab systems. We compare the \textit{exact} $V_{\mathrm{xc}}$ against $n$ and find that there is a high nonlocal dependence on $n$, implying that \textit{no} local density functional can accurately reproduce $V_{\mathrm{xc}}$ and hence $n$ for the slab systems. In light of this, the success of the finite LDAs reported below is all the more surprising.

\section{Testing the LDAs} 
In the previous section we observed the close similarity between the four LDAs. In this section we apply them to a range of model systems \footnote{See Supplemental Material for the parameters of the model systems, and details on our calculations to obtain converged results.} in order to identify the differences between them.

\subsection{Weakly correlated systems}

\begin{figure}[htbp]      
\centering
\includegraphics[width=1.0\linewidth]{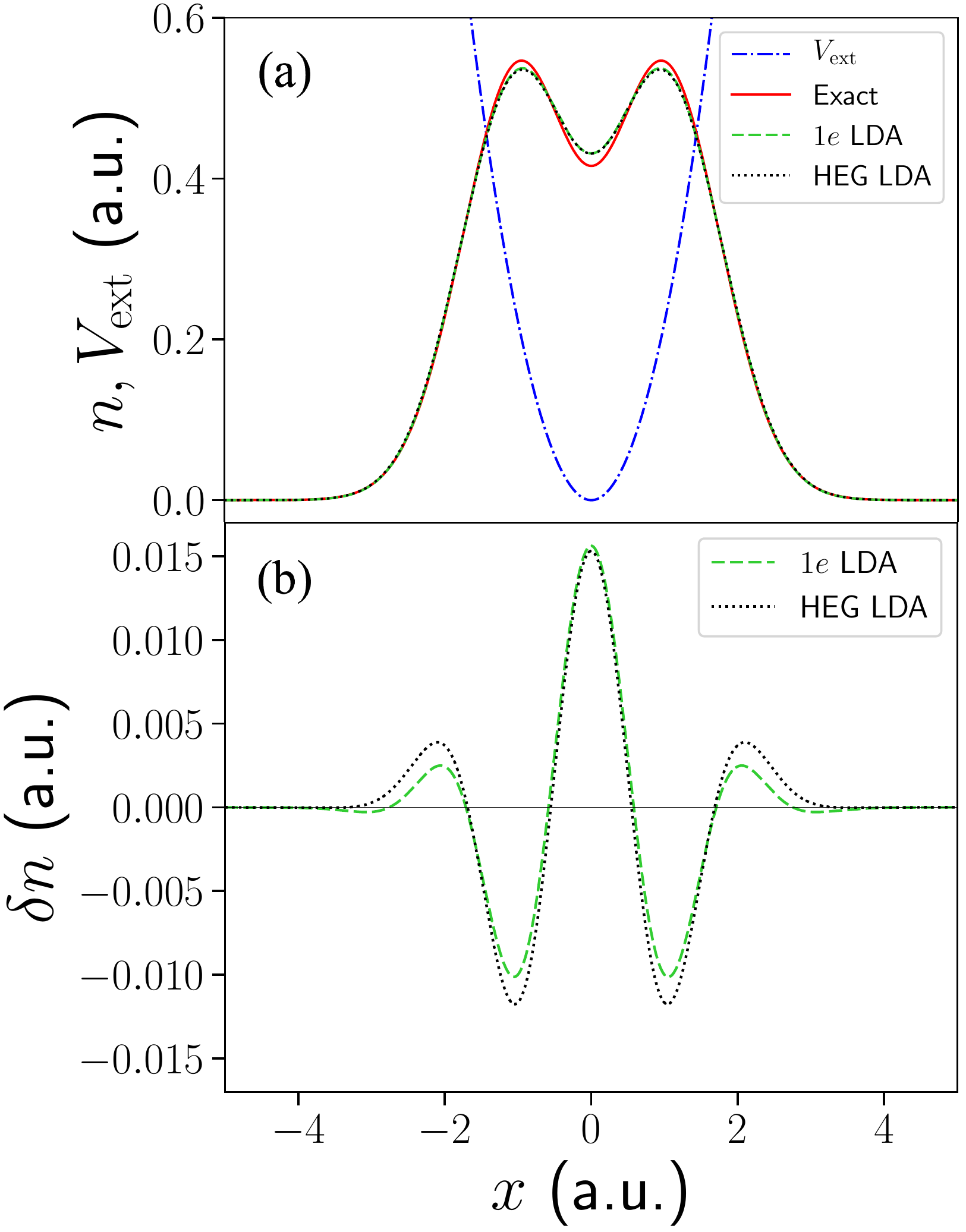}
\caption{System 1 (two electrons in a harmonic potential well). (a) The external potential (dotted-dashed blue line), together with the exact electron density (solid red line), and the densities obtained from applying the $1e$ (dashed green line) and HEG (dotted black line) LDAs. Both LDAs are in very good agreement with the exact result. (b) The absolute error in the density ($\delta n = n^{\mathrm{LDA}} - n^{\mathrm{exact}}$) in the $1e$ (dashed green line) and HEG (dotted black line) LDAs, allowing their differences to be more clearly identified.}
\label{qho2}
\end{figure}

\textit{System 1 (2e harmonic well)}. We first consider a pair of interacting electrons in a strongly confining harmonic potential well ($\omega = \frac{2}{3}$ a.u.) where correlation is very weak \footnote{We calculate the absolute error between the exact electron density and the density obtained from a self-consistent Hartree-Fock calculation ($\delta n = n^{\mathrm{HF}} - n^{\mathrm{exact}}$), and find the net absolute error to be $\int |\delta n| \ dx \approx 1.4 \times 10^{-3}$. The correlation energy is 0.13$\%$ of the exchange-correlation energy, $-0.62$ a.u.}. We calculate the exact many-body electron density using iDEA, and compare it against the densities obtained from applying the LDAs self-consistently. There is a progression from the $1e$--$2e$--$3e$--HEG LDA and so we choose to plot the $1e$ and HEG LDA densities (i.e., the $2e$ and $3e$ LDA densities lie between these) against the exact [Fig.~\ref{qho2}(a)]. Both LDAs match the exact density well, and so we plot their absolute errors ($\delta n = n^{\mathrm{LDA}} - n^{\mathrm{exact}}$) to more clearly identify their differences [Fig.~\ref{qho2}(b)]. The $1e$ LDA has a slightly smaller net absolute error ($\int |\delta n| \ dx$). While the HEG LDA gives a slightly better electron density in the central region (dip in the density), the $1e$ LDA better matches the decay of the density towards the edges of the system, and perhaps more interestingly, the two peaks in the density where the self-interaction correction is largest.

Due to the importance of energies in DFT calculations, we also compare the exact $E_{\mathrm{xc}}$ and total energy $E_{\mathrm{total}}$, with those obtained from applying the LDAs self-consistently (Table~\ref{Exchange_table}). While all the LDAs give good approximations to both quantities, there are some significant differences due to this system being dominated by regions of high density, and the $\varepsilon_{\mathrm{xc}}$ curves separating in this limit (see Fig.~\ref{exc_combined}). As with the approximations to the electron density, there is a progression from the $1e$--$2e$--$3e$--HEG LDA, with the $1e$ LDA reducing the absolute errors ($\frac{\delta E_{\mathrm{xc}}}{E_{\mathrm{xc}}}$, $\frac{\delta E}{E}$) in the HEG LDA by a factor of $5-6$.

\begin{table*}
\caption{\label{Exchange_table}Total energies and xc energies for the set of weakly correlated systems (1--3), from exact calculations and from applying the four LDAs self-consistently ($\delta E^{\mathrm{LDA}} = E^{\mathrm{LDA}} - E^{\mathrm{exact}}$). Estimated errors are $\pm$1 in the last decimal place, unless otherwise stated in parentheses.}
\begin{ruledtabular}
\begin{tabular}{lcccccccccc}
  System                &              &         & $E_{\mathrm{total}}$ (a.u.) &         &             &              &         & $E_{\mathrm{xc}}$ (a.u.) &         &            \B   \\
                                                                                                                                                    \T\B
                        & \text{Exact} & $\delta E_{\mathrm{total}}^{1e}$ & $\delta E_{\mathrm{total}}^{2e}$ & $\delta E_{\mathrm{total}}^{3e}$ & $\delta E_{\mathrm{total}}^{\text{HEG}}$ & \text{Exact} & $\delta E_{\mathrm{xc}}^{1e}$ & $\delta E_{\mathrm{xc}}^{2e}$  & $\delta E_{\mathrm{xc}}^{3e}$ & $\delta E_{\mathrm{xc}}^{\text{HEG}}$ \\
  \hline
  $2e$ harmonic well      & 1.6932       & 0.0037  & 0.0126                      & 0.0153  & 0.0211      & $-0.6192$      & 0.0045  & 0.0137                   & 0.0165  & 0.0225     \T  \\
  $3e$ harmonic well      & 3.1875       & $-0.0073$ & 0.0065                      & 0.0108  & 0.0199      & $-0.9305(5)$      & \ $-0.0058(5)$ &  0.0085(5)                 & 0.0129(5)  & 0.0223(5)         \\
  $2e$ double well        & $-1.0301$      & 0.0237  & 0.0286                      & 0.0296 & 0.0323       & $-0.5349$      & 0.0256  & 0.0317                   & 0.0331  & 0.0363         \\
\end{tabular}
\end{ruledtabular}
\end{table*}

\begin{figure}[htbp]      
\centering
\includegraphics[width=1.0\linewidth]{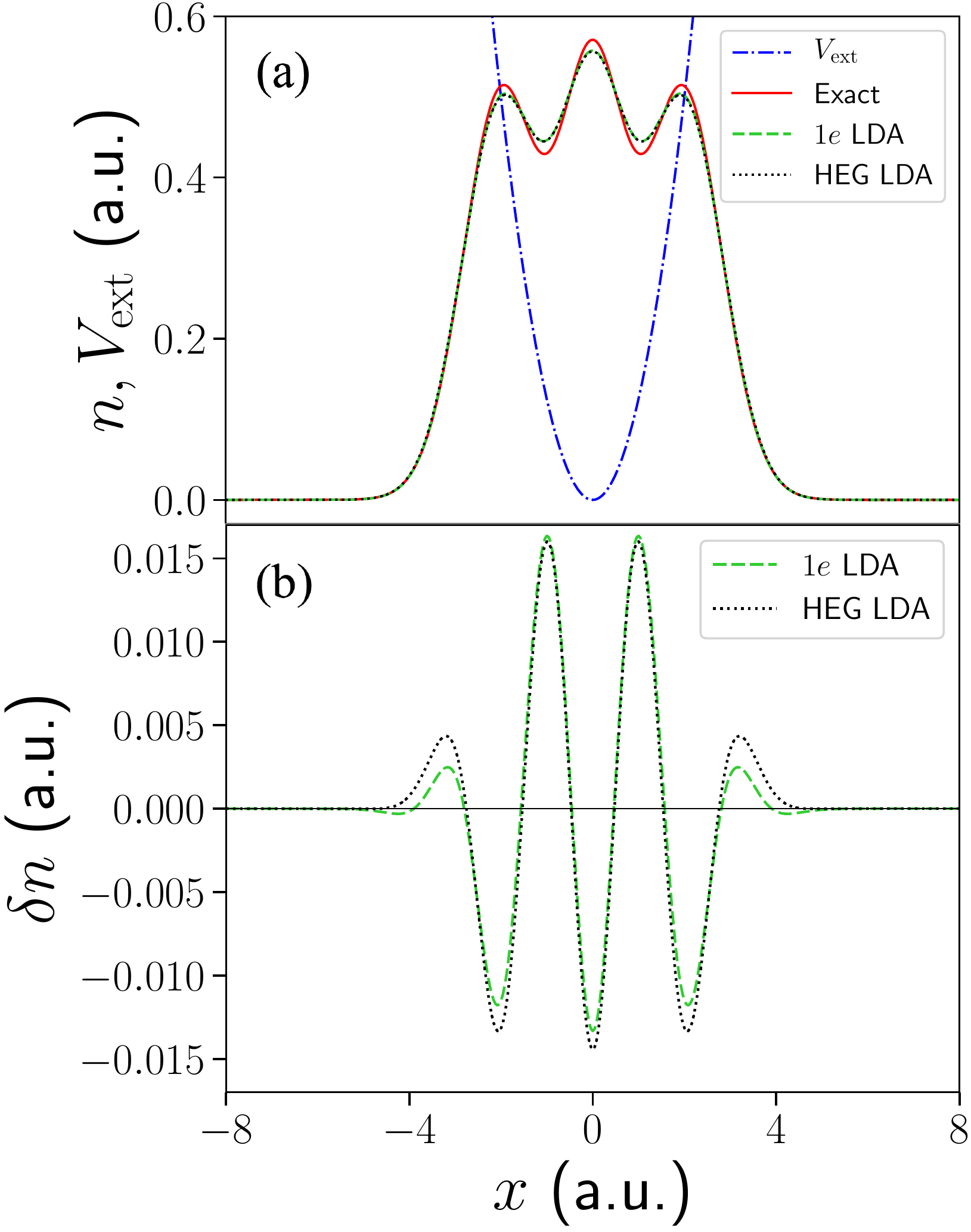}
\caption{System 2 (three electrons in a harmonic potential well). (a) The external potential (dotted-dashed blue line), together with the exact electron density (solid red line), and the densities obtained from applying the $1e$ (dashed green line) and HEG (dotted black line) LDAs. Much like the $2e$ harmonic well system, both LDAs match the exact density well. (b) The absolute error in the density in the $1e$ (dashed green line) and HEG (dotted black line) LDAs. Again, the $1e$ LDA outperforms the HEG LDA in the density peaks, which is dominated by the self-interaction correction.}
\label{qho3}
\end{figure}

\textit{System 2 (3e harmonic well)}. Next, we consider a harmonic potential well with three electrons, but slightly less confining ($\omega = \frac{1}{2}$), in order to avoid an unphysically high electron density ($n > 0.6$ a.u.). As in the $2e$ harmonic well system, we find a progression from the $1e$--$2e$--$3e$--HEG LDA, with all LDAs giving good electron densities (see Fig.~\ref{qho3}(a) for the $1e$ and HEG LDA densities plotted against the exact). Again, the $1e$ LDA has the smallest net absolute error, and outperforms the rest of the LDAs in the regions where the density peaks [Fig.~\ref{qho3}(b)].

We also compare the exact $E_{\mathrm{xc}}$ and $E_{\mathrm{total}}$ against the LDAs (Table~\ref{Exchange_table}). All LDAs give good energies, with some noticeable differences between them due to this system being dominated by regions of high density, like in the $2e$ harmonic well system. However, the \textit{magnitude} of $E_{\mathrm{xc}}$ in the $1e$ LDA is greater than the exact (i.e., it overestimates the amount of exchange $+$ correlation), and subsequently it gives a total energy lower than the exact. While the absolute error in $E_{\mathrm{xc}}$ for each LDA is similar to that in $E_{\mathrm{total}}$, this overestimation of exchange $+$ correlation in the $1e$ LDA results in the $2e$ LDA giving the best total energy.

\subsection{A system dominated by the self-interaction correction} \label{SIC}
 
The self-interaction correction (SIC) is \textit{absent} in xc functionals constructed from the HEG. However, the xc energy of the $1e$ slab systems (which were used to construct the $1e$ LDA) consists entirely of SIC. In the first two model systems, we found that the $1e$ LDA (and indeed the other finite LDAs) better describes the electron density in regions where the SIC is strongest, than the HEG LDA. We now investigate this further. 

\begin{figure}[htbp]      
\centering
\includegraphics[width=1.0\linewidth]{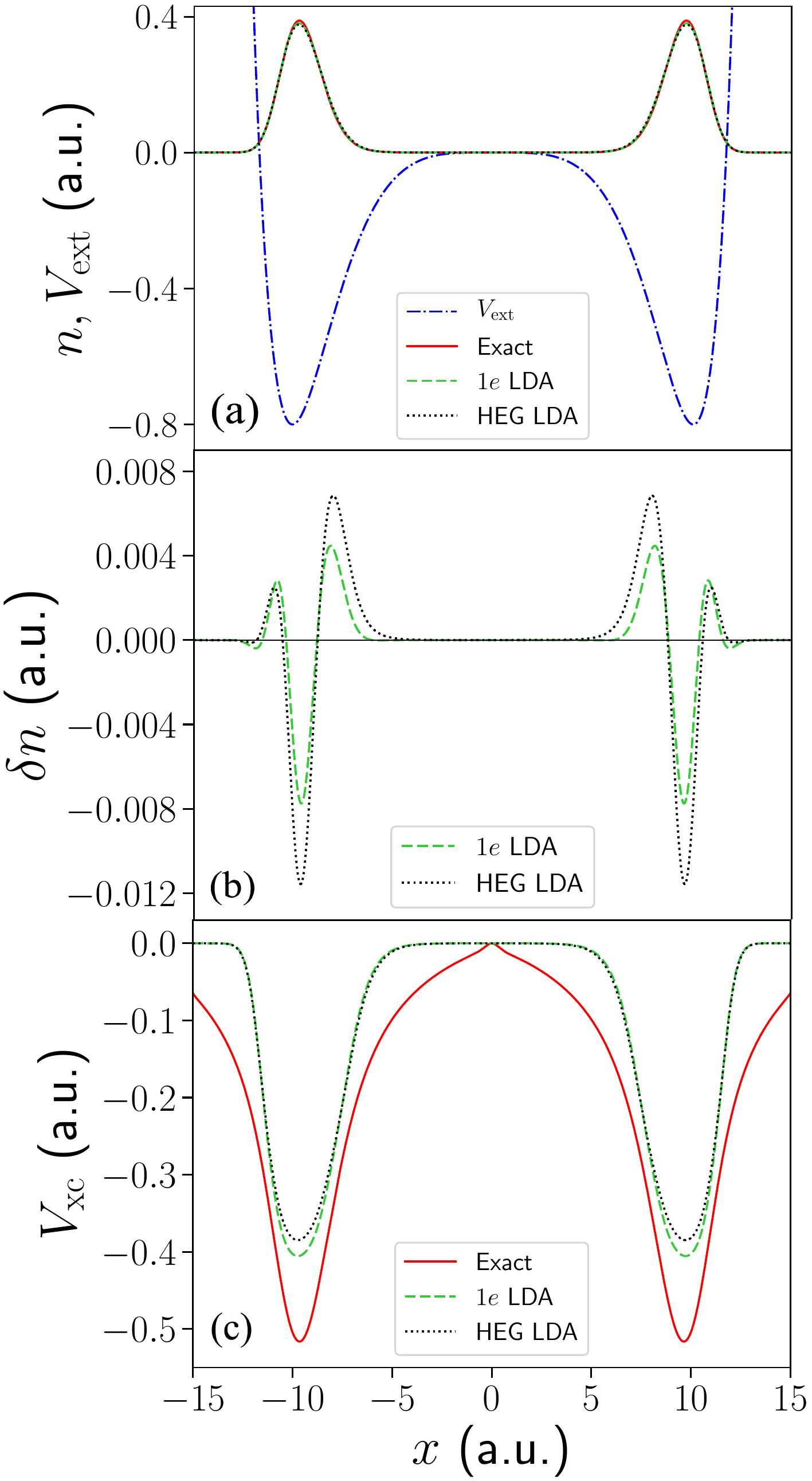}
\caption{System 3 (two electrons in a double-well potential). (a) The external potential (dotted-dashed blue line), together with the exact electron density (solid red line), and the densities obtained from applying the $1e$ (dashed green line) and HEG (dotted black line) LDAs. The wells are separated, such that the electrons are highly localized. (b) The absolute error in the density in the $1e$ (dashed green line) and HEG (dotted black line) LDAs. The $1e$ LDA is far superior in the regions where the density peaks, and hence where the Hartree potential is large and dominated by the electron self-interaction. (c) The exact xc potential (solid red line), and the xc potentials given by the $1e$ (dashed green line) and HEG (dotted black line) LDAs. The dips in $V_{\mathrm{xc}}$ are more closely matched by the $1e$ LDA due to it better capturing the self-interaction correction, present in the exact $V_{\mathrm{xc}}$.}
\label{dw}
\end{figure}

\textit{System 3 (2e double well)}. We choose a system with two electrons confined to a double-well potential. The wells are separated, such that the electrons are \textit{highly localized} and can be considered as two separate subsystems [Fig.~\ref{dw}(a)]. This results in the Hartree potential being small outside of the wells, and being dominated by the electron self-interaction within the wells. Consequently, a large proportion of the xc potential is self-interaction correction. Applying the LDAs, we find the usual progression $1e$--$2e$--$3e$--HEG. Focusing on the peaks in the electron density, the $1e$ LDA substantially reduces the error present in the HEG LDA [Fig.~\ref{dw}(b)]. To understand this, we analyze the xc potential [Fig.~\ref{dw}(c)]. The $1e$ LDA better reproduces the large dips in $V_{\mathrm{xc}}$, corresponding to the peaks in the electron density. Hence, the SIC is more effectively captured.

While the LDA errors in $E_{\mathrm{xc}}$ are larger than in the first two systems, they are still small (4.8--6.8$\%$) (Table~\ref{Exchange_table}). The absolute errors in $E_{\mathrm{total}}$ are similar.

\subsection{Systems where correlation is stronger}

\begin{figure}[htbp]      
\centering
\includegraphics[width=1.0\linewidth]{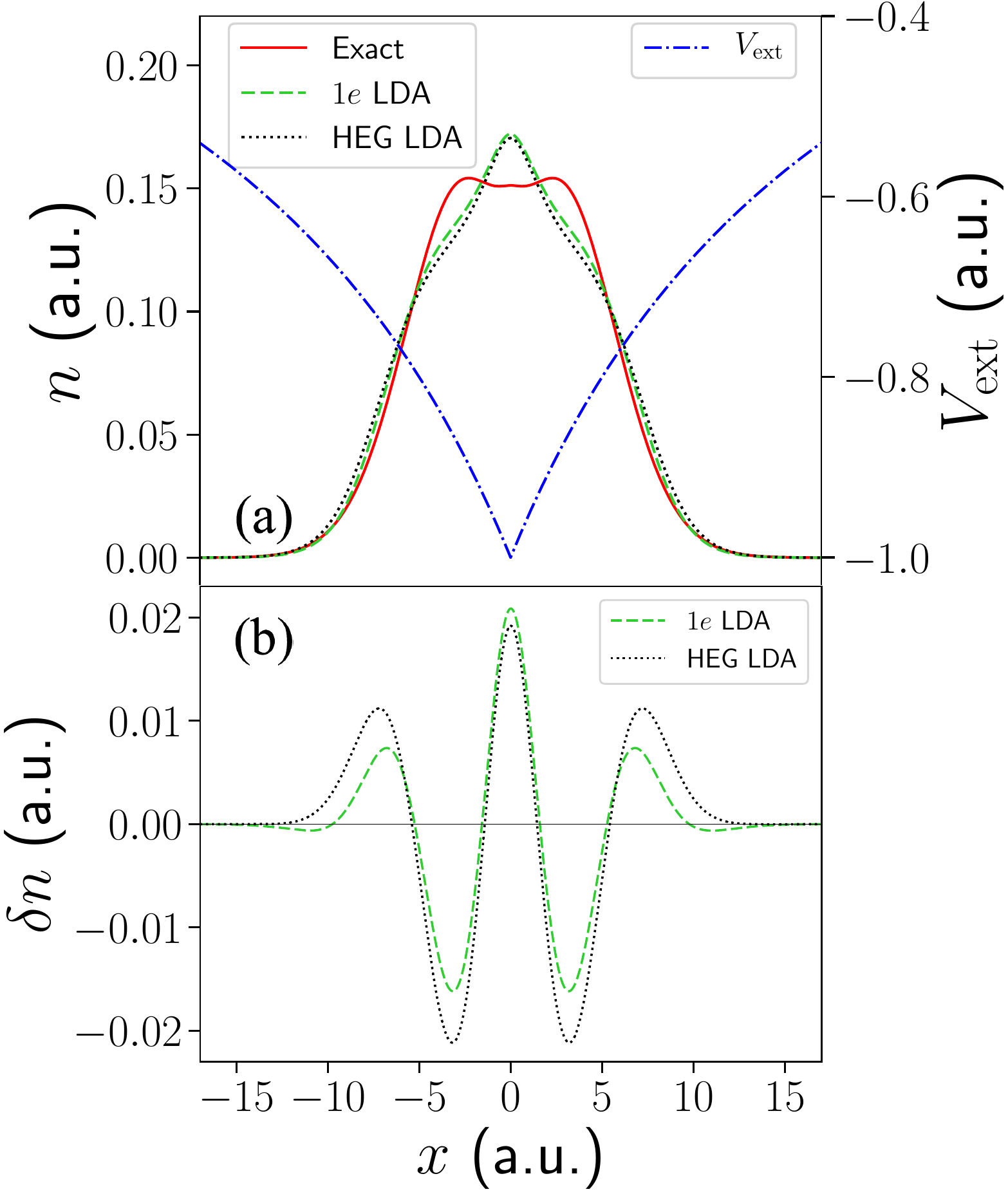}
\caption{System 4 (two electrons in a softened atomiclike potential). (a) The external potential (dotted-dashed blue line), together with the exact electron density (solid red line), and the densities obtained from applying the $1e$ (dashed green line) and HEG (dotted black line) LDAs. Unlike in the weakly correlated systems, the LDAs give poor electron densities. (b) The absolute error in the density in the $1e$ (dashed green line) and HEG (dotted black line) LDAs. While the net absolute errors are much larger than in the weakly correlated systems, the $1e$ LDA still performs the best.}
\label{atom2}
\end{figure}

\textit{System 4 (2e atom)}. We now consider a system where the relative size of electron correlation increases significantly \footnote{We calculate the absolute error between the exact electron density and the density obtained from a self-consistent Hartree-Fock calculation ($\delta n = n^{\mathrm{HF}} - n^{\mathrm{exact}}$), and find the net absolute error to be $\int |\delta n| \ dx \approx 7.4 \times 10^{-2}$. The correlation energy is 1.1$\%$ of the exchange-correlation energy, $-0.37$ a.u.}: two electrons confined to a \textit{softened} atomiclike potential, $V_{\mathrm{ext}} = -(|ax|+1)^{-1}$, where $a = \frac{1}{20}$. Although we find the same progression ($1e$--$2e$--$3e$--HEG) as seen in the first three model systems, in which correlation was weak, all LDAs give inadequate electron densities. This can be seen by plotting the $1e$ and HEG LDA densities against the exact [Fig.~\ref{atom2}(a)]. The LDAs give densities that are not even qualitatively correct, e.g., predicting a single peak in the center of the system, which is absent in the exact density. The net absolute errors are much larger than in the weakly correlated systems, however, the $1e$ LDA once again gives the smallest [Fig.~\ref{atom2}(b)].

We find that although the LDA densities are poor, the xc energies are surprisingly good (Table~\ref{Correlation_table}). This can be attributed \textit{somewhat} (see Sec.~\ref{cancel_errors} for investigation of further causes) to errors in the density being partially canceled by errors inherent in the approximate xc energy functional \cite{Errors_Energy}. We infer this by noting the progression (HEG--$3e$--$2e$--$1e$) when we apply the LDAs to the \textit{exact} density, in contrast to the self-consistent solutions in Table~\ref{Correlation_table}. As in the weakly correlated systems, the absolute errors in $E_{\mathrm{total}}$ are smaller than in $E_{\mathrm{xc}}$, due to a partial cancellation of errors from the Hartree energy component. It is much more apparent in this system due to the LDAs incorrectly predicting a central peak in the electron density [Fig.~\ref{atom2}(a)].      

\begin{table*}
\caption{\label{Correlation_table}Total energies and xc energies for the set of strongly correlated systems (4-5), from exact calculations and from applying the four LDAs self-consistently ($\delta E^{\mathrm{LDA}} = E^{\mathrm{LDA}} - E^{\mathrm{exact}}$). Estimated errors are $\pm$1 in the last decimal place, unless otherwise stated in parentheses.}
\begin{ruledtabular}
\begin{tabular}{lcccccccccc}
  System                &              &         & $E_{\mathrm{total}}$ (a.u.) &         &             &              &         & $E_{\mathrm{xc}}$ (a.u.) &         &            \B   \\
                                                                                 \T\B
                        & \text{Exact} & $\delta E_{\mathrm{total}}^{1e}$ & $\delta E_{\mathrm{total}}^{2e}$ & $\delta E_{\mathrm{total}}^{3e}$ & $\delta E_{\mathrm{total}}^{\text{HEG}}$ & \text{Exact} & $\delta E_{\mathrm{xc}}^{1e}$ & $\delta E_{\mathrm{xc}}^{2e}$  & $\delta E_{\mathrm{xc}}^{3e}$ & $\delta E_{\mathrm{xc}}^{\text{HEG}}$ \\
  \hline
  $2e$ atom      & -1.5099       & 0.0053  & 0.0044                      & 0.0032  & 0.0022      & -0.3728      & 0.0084  & 0.0101                   & 0.0099  & 0.0111     \T  \\
  $3e$ atom      & -2.3282(5)       & 0.0121(5) & 0.0085(5)                      & 0.0057(5)  & 0.0029(5)      & -0.493(4)      & 0.029(4) & 0.029(4)                   & 0.027(4)  & 0.028(4)         \\
\end{tabular}
\end{ruledtabular}
\end{table*}

\begin{figure}[htbp]      
\centering
\includegraphics[width=1.0\linewidth]{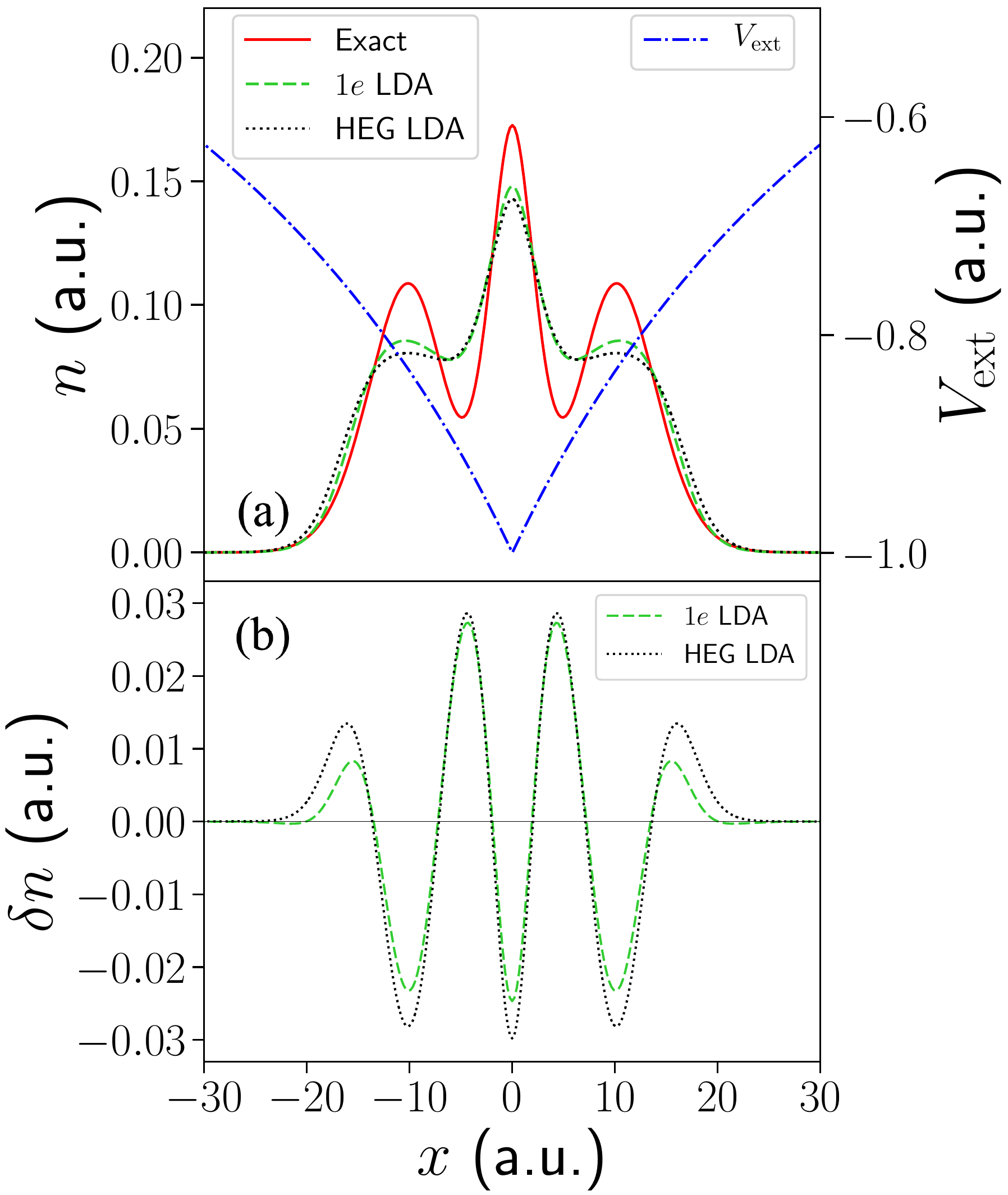}
\caption{System 5 (three electrons in a softened atomiclike potential). (a) The external potential (dotted-dashed blue line), together with the exact electron density (solid red line), and the densities obtained from applying the $1e$ (dashed green line) and HEG (dotted black line) LDAs. Like in the $2e$ atom, the LDAs give poor electron densities. The $1e$ LDA more accurately reproduces the peaks in the density, where the SIC is largest. (b) The absolute error in the density in the $1e$ (dashed green line) and HEG (dotted black line) LDAs. Again, the net absolute errors are large, with the $1e$ LDA giving the smallest.}
\label{atom3}
\end{figure}

\textit{System 5 (3e atom)}. Finally, we consider three electrons in an external potential of the same form as the $2e$ atom, but less confining, with $a = \frac{1}{50}$. Along with the usual progression ($1e$--$2e$--$3e$--HEG), we find a similar result to the $2e$ atom, with the LDAs giving poor electron densities [Fig.~\ref{atom3}(a)]. Although the densities are qualitatively correct, unlike in the $2e$ atom, the LDAs significantly underestimate the peaks in the electron density. Subsequently, the absolute errors are very large [Fig.~\ref{atom3}(b)]. The $1e$ LDA, along with giving the lowest net absolute error, most accurately reproduces the peaks in the density, where the SIC is largest.

While the absolute errors in $E_{\mathrm{xc}}$ are larger than in the $2e$ atom, they are still small (Table~\ref{Correlation_table}). Again, this partially arises from applying approximate xc energy functionals to incorrect densities. As in the $2e$ atom, the absolute errors in $E_{\mathrm{total}}$ are much lower than those in $E_{\mathrm{xc}}$, due to a partial cancellation of errors from the Hartree energy component.

\subsection{Cancellation of errors between exchange and correlation} \label{cancel_errors}

HEG-based LDAs have been known to typically underestimate the \textit{magnitude} of the exchange energy $E_{\mathrm{x}}$, while overestimating the magnitude of the correlation energy $E_{\mathrm{c}}$. Consequently, while the total $E_{\mathrm{xc}}$ is underestimated in magnitude, the approximation proves to be better than was originally expected due to a partial cancellation of errors.

We investigate how well our HEG LDA approximates $E_{\mathrm{x}}$ and $E_{\mathrm{c}}$ in the model systems, and how this contributes to accurate values for $E_{\mathrm{xc}}$. To do this we perform Hartree-Fock calculations for each of the model systems, and together with the exact solutions obtained through iDEA, are able to divide the exact $E_{\mathrm{xc}}$ into its exchange and correlation components. We then apply the HEG LDA, which is split into separate $E_{\mathrm{x}}$ and $E_{\mathrm{c}}$ functionals, for comparison (Table~\ref{xandc_table}). In all systems, the HEG LDA underestimates the magnitude of $E_{\mathrm{x}}$, while it overestimates the magnitude of $E_{\mathrm{c}}$. However, due to the exchange energy being the dominant component of $E_{\mathrm{xc}}$, even in strongly correlated systems, this only leads to a partial cancellation of errors.

The $1e$ LDA yields a larger magnitude for $\varepsilon_{\mathrm{x}}$ than the HEG LDA across the entire density range studied (up to 0.6 a.u.) (Fig.~\ref{ex_comparison}), which arises from a better description of the SIC (Sec.~\ref{SIC}). In the $1e$ LDA correlation is absent. Consequently, the $1e$ xc energies that follow from Tables~\ref{Exchange_table} and ~\ref{Correlation_table} can be considered as approximations to $E_{\mathrm{x}}$. We note that the $1e$ LDA substantially reduces the error in $E_{\mathrm{x}}$ that arises in the HEG LDA \footnote{This is also true in the $2e$ double-well system where correlation is negligible, and the exchange energy is dominated by the SIC.}. We infer that this error reduction will also extend to the $2e$ and $3e$ LDAs. 

\begin{table}
\caption{Exchange energies and correlation energies for all systems (1-5), from exact calculations and from applying the HEG LDA self-consistently ($\delta E^{\mathrm{LDA}} = E^{\mathrm{LDA}} - E^{\mathrm{exact}}$). Estimated errors are $\pm$1 in the last decimal place, unless otherwise stated in parentheses.}
\label{xandc_table}
\resizebox{\columnwidth}{!}{%
\begin{ruledtabular}
\begin{tabular}{l*{3}{>{$}c<{$}}}
  System & \ \ \ \ \ \ \ \ \ \ \ \ \ \ \ \ \ \ \ \ \ \ \ \ \ \ \ \ E_{\mathrm{x}} \ \text{(a.u.)} & \B \\
           \cline{2-3} \T\B
         & \text{Exact} & \delta E_{\mathrm{x}}^{\text{HEG}}        \\
  \hline
  $2e$ harmonic well       & -0.6184       & 0.0268      \T    \\
  $3e$ harmonic well       & -0.9286(5)    & 0.0276(5)             \\
  $2e$ double well         & -0.5349       & 0.0441             \\
  $2e$ atom                & -0.3686       & 0.0185             \\
  $3e$ atom                & -0.488(3)     & 0.041(3) \B \\
  \hline
  System & \ \ \ \ \ \ \ \ \ \ \ \ \ \ \ \ \ \ \ \ \ \ \ \ \ \ \ \ E_{\mathrm{c}} \ \text{(a.u.)} & \T\B \\
           \cline{2-3} \T\B
         & \text{Exact} & \delta E_{\mathrm{c}}^{\text{HEG}}        \\
  \hline
  $2e$ harmonic well       & -0.0008       & -0.0043      \T    \\
  $3e$ harmonic well       & -0.0019       & -0.0053             \\
  $2e$ double well         & -0.0000       & -0.0077             \\
  $2e$ atom                & -0.0042       & -0.0074             \\
  $3e$ atom                & -0.0043(5)    & -0.0142(5) 
\end{tabular}
\end{ruledtabular}
}
\end{table}

\section{Conclusions}
We have constructed an LDA based on the homogeneous electron gas (HEG) through suitable quantum Monte Carlo techniques and find that it is remarkably similar in many regards to a set of three LDAs constructed from finite systems. Applying them to test systems to explore the differences between them, we find that the finite LDAs give better densities and energies in highly confined systems in which correlation is weak. Most interestingly, the LDA constructed from systems of just one electron most accurately describes the self-interaction correction. All LDAs give poor densities in systems where correlation is stronger, but give reasonably good energies, with the HEG LDA giving the best total energies. Across all test systems, the HEG LDA underestimates the magnitude of the exchange energy and overestimates the magnitude of the correlation energy, leading to a partial cancellation of errors. As a consequence of the finite LDAs giving a better description of the self-interaction correction, we infer that they would reduce the error in the exchange energy. Furthermore, we expect that finite LDA functionals will also provide a better treatment of the SIC for spinful electrons. Their derivation and usage could lead to an improved description of the electronic structure in a variety of situations, such as at the onset of Wigner oscillations.

Data created during this research is available by request from the York Research Database \footnote{M. T. Entwistle, M. Casula, and R. W. Godby, (2018), doi:10.15124/65cd1dd8-c240-45a5-9a47-0ac6ff870f51.}.

\begin{acknowledgments}
M.C. acknowledges the GENCI allocation for computer resources under Project No. 0906493. We thank Leopold Talirz for recent developments in the iDEA code, and Matt Hodgson and Jack Wetherell for helpful discussions.
\end{acknowledgments}

\bibliography{Entwistle_2018}
\end{document}


\title{Comparison of local density functionals based on electron gas and finite systems (Supplemental Material)} 
\date{\today}
\author{M.\ T.\ Entwistle}
\affiliation{Department of Physics, University of York, and European Theoretical Spectroscopy Facility, Heslington, York YO10 5DD, United Kingdom}
\author{M. Casula}
\affiliation{Institut de Min\'{e}ralogie, de Physique des Mat\'{e}riaux et de Cosmochimie (IMPMC), Sorbonne Universit\'{e}, CNRS UMR 7590, IRD UMR 206, MNHN, 4 Place Jussieu, 75252 Paris, France}
\author{R.\ W.\ Godby}
\affiliation{Department of Physics, University of York, and European Theoretical Spectroscopy Facility, Heslington, York YO10 5DD, United Kingdom}

\maketitle

\section{Exchange-correlation potentials} 
The xc potential is defined as the functional derivative of the xc energy which in the LDA reduces to a simple form:
\begin{equation}
V_{\mathrm{xc}}^{\mathrm{LDA}}(n) = \frac{\delta E_{\mathrm{xc}}^{\mathrm{LDA}}[n]}{\delta n(x)} = \varepsilon_{\mathrm{xc}}(n(x)) + n(x)\frac{d\varepsilon_{\mathrm{xc}}}{dn}\bigg|_{n(x)}. \label{V_xc} 
\end{equation}

\subsection{Finite LDAs}
The xc energy density in the three finite LDAs is parameterized by (we use Hartree atomic units):
\begin{equation} 
\varepsilon_{\mathrm{xc}}(n) = (A + B n + C n^{2} + D n^{3} + E n^{4} + F n^{5}) n^{G}. \label{exc_finite}
\end{equation}

The xc potential is obtained using Eq.~(\ref{V_xc}) :
\begin{equation}
\begin{split}
V_{\mathrm{xc}}^{\mathrm{LDA}}(n) & =  \big[A(1+G) + B(2+G)n + C(3+G)n^{2} \\ 
& \ \ \ \ + D(4+G)n^{3} + E(5+G)n^{4} \\
& \ \ \ \ + F(6+G)n^{5}\big]n^{G}. \label{vxc_finite}
\end{split}
\end{equation}

\subsection{HEG LDA}
In the HEG LDA, the xc energy density is split into separate exchange and correlation parts. Consequently, Eq.~(\ref{V_xc}) is also split into separate exchange and correlation parts.
\\ \\ 
The exchange energy density in the HEG LDA is parameterized as in Eq.~(\ref{exc_finite}), and so the exchange potential is of the same form as Eq.~(\ref{vxc_finite}). 
\\ \\
The correlation energy density in the HEG LDA is parameterized by:
\begin{equation}
\varepsilon_{\mathrm{c}}(r_{\mathrm{s}}) = -\frac{A_{\mathrm{RPA}} r_{\mathrm{s}} + E r_{\mathrm{s}}^{2}}{1 + B r_{\mathrm{s}} + C r_{\mathrm{s}}^{2} + D r_{\mathrm{s}}^{3}} \frac{\ln(1 + \alpha r_{\mathrm{s}} + \beta r_{\mathrm{s}}^{2})}{\alpha}, \label{ec_heg}
\end{equation}
where $2 r_{\mathrm{s}} = 1/n$. The correlation potential is given by:
\begin{equation}
\begin{split}
V_{\mathrm{c}}(r_{\mathrm{s}}) = \varepsilon_{\mathrm{c}}(r_{\mathrm{s}}) \ & - \frac{r_{\mathrm{s}}}{\alpha (1+B r_{\mathrm{s}}+C r_{\mathrm{s}}^{2}+D r_{\mathrm{s}}^{3})} \Bigg[ -(A + 2 E r_{\mathrm{s}}) \ln(1 + \alpha r_{\mathrm{s}} + \beta r_{\mathrm{s}}^{2})  \\
& +\frac{(A r_{\mathrm{s}} + E r_{\mathrm{s}}^{2})(B+2 C r_{\mathrm{s}} + 3 D r_{\mathrm{s}}^{2}) \ln(1 + \alpha r_{\mathrm{s}} + \beta r_{\mathrm{s}}^{2})}{1 + B r_{\mathrm{s}} + C r_{\mathrm{s}}^{2} + D r_{\mathrm{s}}^{3}} -\frac{(A r_{\mathrm{s}} + E r_{\mathrm{s}}^{2})(\alpha + 2 \beta r_{\mathrm{s}})}{(1 + \alpha r_{\mathrm{s}} + \beta r_{\mathrm{s}}^{2})} \Bigg].
\end{split}
\end{equation}

\section{System 1 (Two-electron harmonic well)}
The external potential is:
\begin{equation} \label{qho_potential}
V_{\mathrm{ext}}(x) = \frac{1}{2} \omega^{2} x^{2},
\end{equation}
where $\omega = \frac{2}{3}$ a.u. For this system converged results are obtained with $\delta x = 0.01$ a.u. for exact and Hartree-Fock calculations, and $\delta x = 0.005$ a.u. for LDA calculations.

\section{System 2 (Three-electron harmonic well)}
The external potential is of the same form as Eq.~(\ref{qho_potential}) with $\omega = \frac{1}{2}$ a.u. The grid spacing is $\delta x = 0.04$ a.u. for exact and Hartree-Fock calculations, and $\delta x = 0.004$ a.u. for LDA calculations.

\section{System 3 (Two-electron double well)}
The external potential is:
\begin{equation} 
V_{\mathrm{ext}}(x) = \alpha x^{10} - \beta x^{4},
\end{equation}
where $\alpha = 5 \times 10^{-11}$ and $\beta = 1.3 \times 10^{-4}$. The grid spacing is $\delta x = 0.015$ a.u. for exact and Hartree-Fock calculations, and $\delta x = 0.0075$ a.u. for LDA calculations.

\section{System 4 (Two-electron atom)}
The external potential is:
\begin{equation} \label{atom_potential}
V_{\mathrm{ext}}(x) = -\frac{1}{|ax|+1},
\end{equation}
where $a = \frac{1}{20}$. The grid spacing is $\delta x = 0.03$ a.u. for exact and Hartree-Fock calculations, and $\delta x = 0.015$ a.u. for LDA calculations.

\section{System 5 (Three-electron atom)}
The external potential is of the same form as Eq.~(\ref{atom_potential}) with $a = \frac{1}{50}$ a.u. The grid spacing is $\delta x = 0.225$ a.u. for exact and Hartree-Fock calculations, and $\delta x = 0.0225$ a.u. for LDA calculations.